\documentclass[twocolumn]{aastex631}

\usepackage{amsmath}

\begin{document}

\title{Accurately Estimating Redshifts from CSST Slitless Spectroscopic Survey using Deep Learning}

\correspondingauthor{Yan Gong}
\email{gongyan@bao.ac.cn}

\author{Xingchen Zhou}
\affiliation{National Astronomical Observatories, Chinese Academy of Sciences, 20A Datun Road, Beijing 100101, People's Republic of China}
\affiliation{Science Center for China Space Station Telescope, National Astronomical Observatories, Chinese Academy of Sciences, 20A Datun Road, Beijing 100101, People's Republic of China}

\author[0000-0003-0709-0101]{Yan Gong}
\affiliation{National Astronomical Observatories, Chinese Academy of Sciences, 20A Datun Road, Beijing 100101, People's Republic of China}
\affiliation{University of Chinese Academy of Sciences, Beijing, 100049, People's Republic of China}
\affiliation{Science Center for China Space Station Telescope, National Astronomical Observatories, Chinese Academy of Sciences, 20A Datun Road, Beijing 100101, People's Republic of China}

\author{Xin Zhang}
\affiliation{National Astronomical Observatories, Chinese Academy of Sciences, 20A Datun Road, Beijing 100101, People's Republic of China}
\affiliation{Science Center for China Space Station Telescope, National Astronomical Observatories, Chinese Academy of Sciences, 20A Datun Road, Beijing 100101, People's Republic of China}

\author{Nan Li}
\affiliation{National Astronomical Observatories, Chinese Academy of Sciences, 20A Datun Road, Beijing 100101, People's Republic of China}
\affiliation{Science Center for China Space Station Telescope, National Astronomical Observatories, Chinese Academy of Sciences, 20A Datun Road, Beijing 100101, People's Republic of China}

\author{Xian-Min Meng}
\affiliation{National Astronomical Observatories, Chinese Academy of Sciences, 20A Datun Road, Beijing 100101, People's Republic of China}
\affiliation{Science Center for China Space Station Telescope, National Astronomical Observatories, Chinese Academy of Sciences, 20A Datun Road, Beijing 100101, People's Republic of China}

\author{Xuelei Chen}
\affiliation{National Astronomical Observatories, Chinese Academy of Sciences, 20A Datun Road, Beijing 100101, People's Republic of China}
\affiliation{Science Center for China Space Station Telescope, National Astronomical Observatories, Chinese Academy of Sciences, 20A Datun Road, Beijing 100101, People's Republic of China}
\affiliation{Centre for High Energy Physiscs, Peking University, Beijing 100871, Peoples Republic of China}

\author{Run Wen}
\affiliation{Purple Mountain Observatory, Chinese Academy of Sciences, 10 Yuanhua Road, Nanjing 210023, People's Republic of China}
\affiliation{School of Astronomy and Space Sciences, University of Science and Technology of China, Hefei, 230026, People's Republic of China}

\author{Yunkun Han}
\affiliation{Yunnan Observatories, Chinese Academy of Sciences, 396 Yangfangwang, Guandu District, Kunming 650216, People's Republic of China}

\author{Hu Zou}
\affiliation{National Astronomical Observatories, Chinese Academy of Sciences, 20A Datun Road, Beijing 100101, People's Republic of China}

\author{Xian Zhong Zheng}
\affiliation{Purple Mountain Observatory, Chinese Academy of Sciences, 10 Yuanhua Road, Nanjing 210023, People's Republic of China}
\affiliation{School of Astronomy and Space Sciences, University of Science and Technology of China, Hefei, 230026, People's Republic of China}

\author{Xiaohu Yang}
\affiliation{Department of Astronomy, School of Physics and Astronomy, Shanghai Jiao Tong University, Shanghai 200240, People's Republic of China}
\affiliation{Tsung-Dao Lee Institute and Key Laboratory for Particle Physics, Astrophysics and Cosmology, Ministry of Education, Shanghai 201210, People's Republic of China}

\author{Hong Guo}
\affiliation{Shanghai Astronomical Observatory, Chinese Academy of Sciences, Shanghai 200030, People's Republic of China}

\author{Pengjie Zhang}
\affiliation{Department of Astronomy, School of Physics and Astronomy, Shanghai Jiao Tong University, Shanghai 200240, People's Republic of China}
\affiliation{Tsung-Dao Lee Institute and Key Laboratory for Particle Physics, Astrophysics and Cosmology, Ministry of Education, Shanghai 201210, People's Republic of China}




\begin{abstract}

Chinese Space Station Telescope (CSST) has the capability to conduct slitless spectroscopic survey simultaneously with photometric survey. The spectroscopic survey will measure slitless spectra, potentially providing more accurate estimations of galaxy properties, particularly redshifts, compared to using broadband photometry. {CSST relies on these accurate redshifts to perform baryon acoustic oscilliation (BAO) and other probes to constrain the cosmological parameters.} {However, due to low resolution and signal-to-noise ratio of slitless spectra, measurement of redshifts is significantly challenging.} In this study, we employ a Bayesian neural network (BNN) to assess the accuracy of redshift estimations from slitless spectra anticipated to be observed by CSST. The simulation of slitless spectra is based on real observational data from the early data release of the Dark Energy Spectroscopic Instrument (DESI-EDR) and the 16th data release of the Baryon Oscillation Spectroscopic Survey (BOSS-DR16), combined with the 9th data release of the DESI Legacy Survey (DESI LS DR9). {The BNN is constructed employing transfer learning technique, by appending two Bayesian layers after a convolutional neural network (CNN), leveraging the features learned from the slitless spectra and corresponding redshifts.} Our network can provide redshift estimates along with corresponding uncertainties, achieving an accuracy of $\sigma_{\rm NMAD} = 0.00063$, outlier percentage $\eta=0.92\%$ and weighted mean uncertainty $\overline{E} = 0.00228$. {These results successfully fulfill the requirement of $\sigma_{\rm NMAD} < 0.005$ for BAO and other studies employing CSST slitless spectroscopic surveys.}
\end{abstract}



\section{Introduction} \label{sec:intro}
Redshifts are one of the fundamental quantities for studying galaxies. The most accurate redshifts are determined through observing and analyzing high-resolution spectra from galaixes. However, obtaining high-resolution spectra is a time-consuming task, especially for high-redshift and faint sources, which require hours of observation to successfully measure their redshifts. As a result, photometric redshifts, estimated from several photometric mesurements, become a necessary option for most sources observed in ongoing and future cosmological surveys. {However, the best accuracy of $\sigma_{\rm NMAD}$ can barely achieve $\sim0.01$ among photometric redshift estimation endeavours utilizing real observational data~\citep{Zou2019,Schuldt2021,Treyer2024,Jones2024}}. This accuracy severely hinders certain cosmological studies using techniques such as baryon acoustic oscillation (BAO, ~\citet{Bassett2010}) and redshift-space distortions (RSD, ~\citet{Hamilton1998}). To match the accuracy required by these cosmological studies and the survey speed of current photometric surveys, a compromise solution exists: slitless spectra. Slitless spectra represent a category of low-resolution spectroscopy performed without a narrow slit, which typically allows only light from a small region to be diffracted. Current and future photometric surveys, such as Euclid Space Telescope (Euclid, ~\citet{Euclidcoll2024}), James Webb Space Telescope (JWST, ~\citet{Sabelhaus2004}), Nancy Grace Roman Space Telescope (Roman)~\footnote{\url{https://roman.gsfc.nasa.gov/}} and Chinese Space Station Telescope (CSST, ~\citet{Zhan2018, Gong2019}), all include modules to observe slitless spectra for galaxies. 

CSST is a 2-meter space telescope designed for photometric survey across seven bands, ranging from near-ultraviolet to near-infrared. The slitless spectroscopy module, which include three bands ($GU$, $GV$ and $GI$), operates alongside the photometric module, enabling simultaneous photometric and slitless spectroscopic observations. These three bands can reach 5$\sigma$ magnitude limit of 23.2, 23.4 and 23.2 for point sources, respectively, with a low spectroscopic resolution of each band as $R=\lambda/\Delta\lambda\geq 200$~\citep{Gong2019}. For extended sources such as galaxies, slitless spectra can be significantly affected by observational and instrumental effects, challenging the one-dimensional spectrum extraction procedure and thus resulting in low-resolution and signal-to-noise ratio (SNR) spectra. {These challenges bring significant difficulties in the recognition of emission and absorption lines, breaks, and other spectroscopic features.} As a result, galaxy properties such as redshift and line fluxes estimated from these spectra can be highly inaccurate, {leading to low accuracy even comparable to those derived from broadband photometry.} Addressing the challenge of successfully measuring these galaxy properties from such low-resolution and SNR slitless spectra remains an urgent problem. 

Machine Learning (ML), particularly Deep Learning (DL) algorithms (also known as neural networks), offers a potential solution to the challenges. This algorithm can effectively learn the inherent correlations between inputs and outputs using large datasets, making them well-suited to handle data significantly affected by instrumental or other forms of noise. In the astronomical and cosmological communities, neural networks have gained prominence in recent years, achieving applications across various fields. The multilayer perceptron (MLP), a simple neural network, has been applied to estimate photometric redshifts from multi-band photometric measurements~\citep{Collister2004,Sadeh2014,Zhou2022}, surpassing the accuracy achieved by traditional spectral energy distribution (SED) fitting methods. 

Furthermore, state-of-the-art convolutional neural networks (CNN, ~\citet{Lecun1995}), which excel in directly processing images, have become indispensable in astronomical and cosmological analysis. Applications include deriving photometric redshifts or other quantities from galaxy images~\citep{Pasquet2019, Henghes2022, Zhou2022, Tewes2019, Zhang2024}, discovering strong lensing systems or mergers~\citep{He2020, Schaefer2018, Li2020, Rezaei2022, Pearson2019, Arendt2024}, and constraining cosmological parameters from large-scale structures or weak gravitational lensing~\citep{Pan2020,Min2024,hortua2023,Gupta2018,Fluri2022}.

In addition to processing 2-dimensional arrays, CNNs can be adapted to handle 1-dimensional sequences or 3-dimensional data cubes. Spectra are 1-dimensional sequences containing redshift or other information, which can be effectively extracted by 1d-CNNs. The application of 1d-CNNs for deriving redshifts from spectra is extensively researched~\citep{Rastegarnia2022, Busca2018}. 

Unlike traditional fitting methods that produce both redshifts and uncertainties, deep learning methods typically provide only redshift values. {Recognizing the importance of uncertainties in cosmological studies, Bayesian neural networks (BNN)~\citep{MacKay1995,Blundell2015,Gal2015}, which can output both point estimations and uncertainties, have gained significant attention, especially in cosmological studies~\citep{Perrault2017, Hortua2020, Zhou2022b, hortua2023, Jones2024}.} By assigning probability distributions to each weight in the network, BNNs can capture and propagate uncertainties from data and neural network itself to the output, providing not only point predictions but also confidence intervals or posterior distributions. 

Although deep learning algorithm offers advantages in providing better accuracy, higher speed, and direct processing of raw data, several challenges need careful consideration. Since deep learning models heavily rely on training data, obtaining abundant and representative data for observation is a primary concern. Specifically, for redshift estimation, a large dataset with high-quality spectroscopic redshifts is essential. Fortunately, several ongoing and planned spetroscopic surveys, such as the Dark Energy Spectroscopic Instrument (DESI, \citet{DESIcoll2016}), Prime Focus Spectrograph (PFS, ~\citet{Tamura2016}), MUltiplexed Spectroscopic Telescope (MUST)~\footnote{\url{https://must.astro.tsinghua.edu.cn/en}}, MegaMapper~\citep{Schlegel2022} and Wide-field Spectroscopic Telescope (WST, \citet{Mainieri2024}), aim to observe a substantial number of galaxy spectra with accurate redshifts. Including completed surveys like zCOSMOS~\citep{Lilly2007}, VIMOS-VLT Deep Survey (VVDS, \citet{LeFevre2013}), Sloan Digital Sky Survey (SDSS, ~\citet{Ahumada2020}), Baryon Oscillation Spectroscopic Survey (BOSS, ~\citet{Dawson2013}), a sufficient and representative training set for redshift estimation can be achieved. 

{In this work, we employ deep learning technique to estimate spectroscopic redshifts for slitless spectra expected to be observed by CSST. The slitless spectra are simulated based on real spectroscopic observations.} Considering the redshift coverage and survey fields, we employ data from DESI early data release (DESI-EDR, \citet{DESIcoll2023}) and BOSS 16th data release (BOSS-DR16, \citet{Dawson2013}), combined with 9th data release of DESI Legacy Surveys (DESI LS DR9, \citet{Schlegel2021}). DESI-EDR has made available 1.2 million high-resolution spectra of galaxies and quasars collected during the Survey Validation (SV) phase for target selections. Since the number of sources in DESI-EDR are limited, we supplement our slitless spectrum dataset with BOSS data, which shares a similar pipeline for the measurement of spectroscopic redshifts, to increase the data size for training the neural network model. After obtaining the slitless spectra, we train a 1d-BNN with these spectra and their corresponding accurate spectroscopic redshifts, and analyze the accuracy that can be achieved for CSST slitless spectroscopic survey.

The structure of this paper is organized as follows: we briefly describe CSST slitless spectra simulation software and then explain the generation of the mock slitless spectra in Section~\ref{sec:mock data}. And the neural network methods including CNN and BNN are introduced in Section~\ref{sec:methods}. Then we demonstrate our results in Section~\ref{sec:results}. {The limitations of current study are extensively discussed in Section~\ref{sec:discussion}.} Finally, this paper is concluded in Section~\ref{sec:conclusion}. 

\section{Mock Data}\label{sec:mock data}
In this section, we firstly introduce the slitless spectra simulation software in CSST data analysis pipeline, and then explain the data generation procedure of slitless spectra using this software from real spectroscopic observations. 

\subsection{Slitless spectra simulation software}\label{sec:software}
The simulation software for slitless spectra is an integral part of the CSST data analysis pipeline, with the code available online\footnote{\url{https://csst-tb.bao.ac.cn/code/zhangxin/sls_1d_spec}}. We provide a brief overview of the workflow here, and interested readers are recommended to consult Zhang et al. (in preparation) for detailed information. This software utilizes spectral energy distributions (SEDs) and morphological parameters of galaxies to generate mock spectra. Initially, the dispersion curve for the grating is determined through a fitting process that considers the spectroscopic properties of the CSST's slitless spectrum. {Following this, the energy profile of the galaxy, assumed in S{\'e}rsic profile~\citep{Sersic1963, Sersic1968} derived from its morphological parameters, is converted into a pixelated galaxy image.} Each pixel of the galaxy image undergoes dispersion based on the dispersion curve specific to the CSST grating, in conjunction with the sensitivity curve of the CSST instrument and the galaxy's SED. Finaly, all dispersed components are integrated into a two-dimensional slitless spectral image. Additionally, instrumental effects are simulated using a point spread function (PSF), assumed to be 2D Gaussian distribution with a full width at half maximum (FWHM) of 0.3$^{\prime\prime}$. The sky backgrounds, including zodiacal and earthshine components, are computed as 0.019, 0.214, 0.329 $\rm e^- s^{-1} pixel^{-1}$ for the $GU$, $GV$ and $GI$ bands respectively. To mitigate the effects of instrumental and background noise, we co-add spectra from four exposures, each with a duration of 150 seconds. Following these procedures, we generate first-order spectral images expected to be observed by CSST, from which the 1d-spectra and the corresponding errors can be extracted. 

\subsection{Data generation}\label{sec: data generation}
To realistically simulate our slitless spectra, we utilize spectroscopic observations from the Dark Energy Spectroscopic Instrument (DESI) and the Baryon Oscillation Spectroscopic Survey (BOSS). DESI is an ongoing spectroscopic survey conducted on the Mayall 4-meter telescope at Kitt Peak National Observatory. Over its 5-year mission, DESI aims to observe spectra for more than 30 million galaxies and quasars across 14,000 square degrees of sky~\citep{DESIcoll2016}. Recently, DESI has released its Early Data Release (EDR), which includes spectroscopic data for 1.8 million targets observed during the Survey Validation (SV) phase conducted from December 2020 to June 2021~\citep{DESIcoll2023}. 

\begin{figure}
    \centering
    \includegraphics[width=0.5\textwidth]{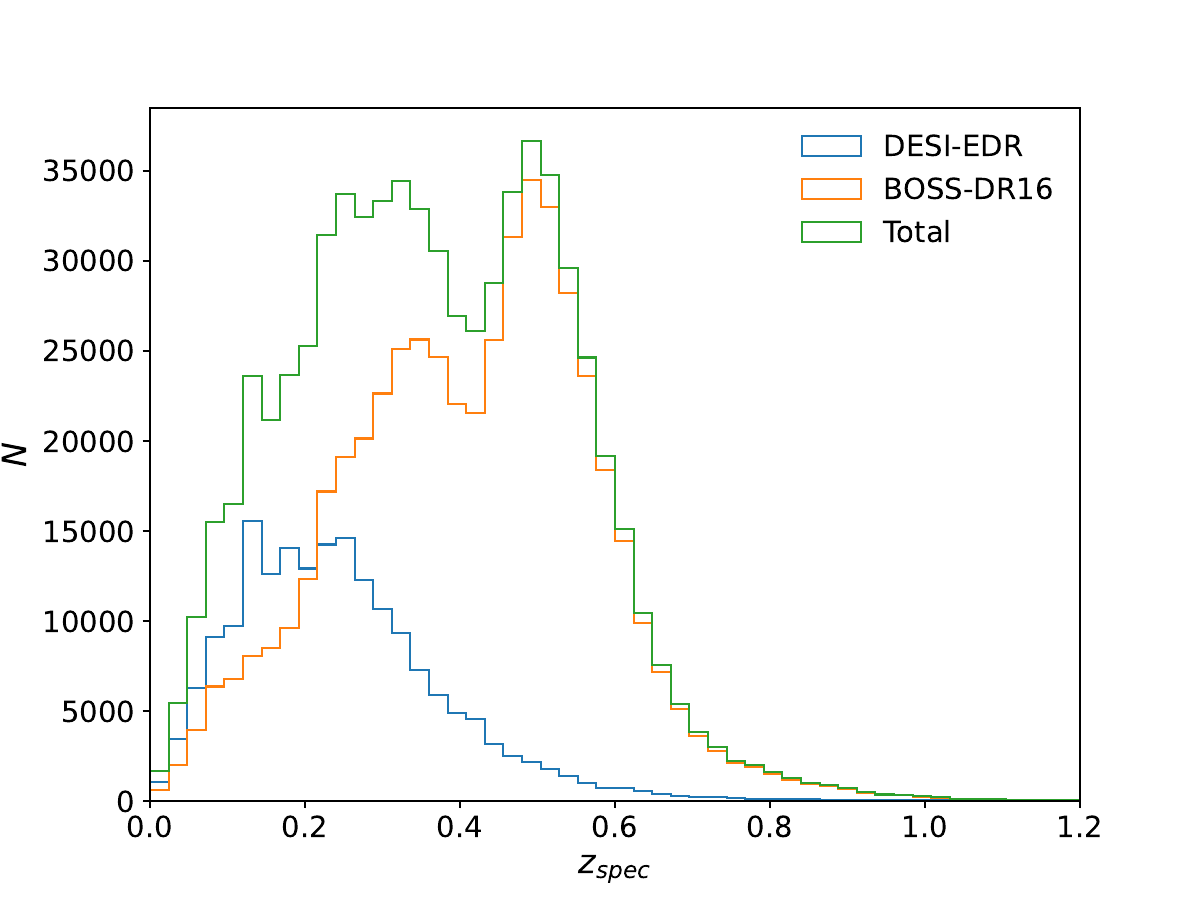}
    \caption{Spectroscopic redshift distributions for selected sources in DESI-EDR, BOSS-DR16 and total.}
    \label{fig:z_dist}
\end{figure}

We select sources from the EDR spectroscopic redshift catalogue using the following criteria:
\begin{equation}
    \begin{aligned}
        &\rm SV\_PRIMARY == True\\
        &\rm MASKBITS == 0\\
        &\rm SPECTYPE == GALAXY\\
        &\rm ZWARN == 0\\
        &\rm FLUX\_G, R, Z > 0 \\
        &\rm FLUX\_IVAR\_G, R, Z > 0 \\
        &\rm MORPHTYPE\ != PSF
    \end{aligned}
\end{equation}
Here, SV\_PRIMARY indicates the best recommended redshift if the same source appears multiple times in the catalogue, while MASKBITS is the bitwise mask indicating that the source touches a pixel in a masked area. SPECTYPE and ZWARN are source classification and indicators of potential issues in spectroscopic redshift measured by Redrock~\footnote{\url{https://github.com/desihub/redrock}}, a commonly used redshift fitting software. We further control the quality of sources by applying constraints on photometric measurements in the $g$, $r$ and $z$ bands of the DESI legacy imaging survey~\citep{Dey2019}. MORPHTYPE indicates the Tractor model used to fit the source during photometric measurement. This constraint ensures that the selected sources are extended, allowing for accurate morphological parameter measurements. It should be noted that some PSF sources are spectroscopically classified as galaxies. These PSF models are assigned probably due to the resolution of $\sim1.0^{\prime\prime}$ in imaging data of DESI legacy survey, and we simply exclude these galaxies in our dataset. To obtain the morphological parameters required to derive our slitless spectra, we match the selected sources with the sweep catalogue of DESI legacy survey DR9~\footnote{\url{https://www.legacysurvey.org/dr9/files/\#sweep-catalogs-region-sweep}} and retrieve the morphological parameters, including effective radius $r_{\rm eff}$, S{\'e}rsic index $n$, two ellipticity components $\epsilon_1, \epsilon_2$ and their variance. And we perform another selection to filter the sources with valid morphological measurements:
\begin{equation}
    \begin{aligned}
        &\rm SHAPE\_R > 0\\
        &\rm SHAPE\_IVAR\_R > 0\\
        &\rm SHAPE\_E1\_IVAR > 0 \\
        &\rm SHAPE\_E2\_IVAR > 0\\
        &\rm SERSIC\_IVAR > 0 
    \end{aligned}
\end{equation}
and then we calculate the axis ratio $b/a$ and position angle $\phi$ using the following equations as recommended by DESI~\footnote{\url{https://www.legacysurvey.org/dr10/catalogs/\#ellipticities}}:
\begin{equation}
    \begin{aligned}
        |\epsilon| = \sqrt{\epsilon_1^2 + \epsilon_2^2}, \\
        \frac{b}{a} = \frac{1 - |\epsilon|}{1 + |\epsilon|}, \\ 
        \phi = \frac{1}{2}\arctan{\frac{\epsilon_2}{\epsilon_1}}.
    \end{aligned}
\end{equation}
{This selection process results in approximately 180,000 sources with high-quality spectroscopic redshifts, with distribution illustrated in blue histogram in Figure~\ref{fig:z_dist}.} {The majority of these sources are bright galaxy samples (BGS) and luminous redshift galaxies (LRG) in DESI, since these categories of sources are larger in size, and consequently, yield measurable morphological parameters. It is acknowledged that the morphological parameters of certain galaxies exhibit significant uncertainties. Furthermore, the morphological variations of some galaxies in spatial dimensions and across different spectral bands cannot be fully captured by single S{\'e}rsic profiles, thereby diminishing the realness of simulated slitless spectra. These effects are beyond the scope of this study and will be addressed in future research.}

After obtaining the redshifts and morphological parameters, the next step involves acquiring the spectral energy distribution (SED) for each source to simulate slitless spectra. The spectroscopic redshifts of the sources have been determined using model spectra fitted by Redrock, and all redshift warning flags are zero, indicating no issues with the fitting process. This allows us to use the model spectra to accurately represent the SED of each source. These model spectra can be constructed using the COEFF provided in the DESI-EDR catalogue, combining Redrock templates, or accessed via the SPectra Analysis \& Retrievable Catalog Lab (SPARCL)~\footnote{\url{https://astrosparcl.datalab.noirlab.edu/}}. The model spectra obtained through both approaches are the same, and we choose the latter. 

{The 180,000 sources selected from DESI-EDR are insufficient for investigating the redshift accuracy in CSST slitless spectroscopic survey, since the redshift distribution displayed in Figure~\ref{fig:z_dist} can barely reach 0.8, much lower than the estimated redshift limit of about 1.5 expected to be observed by CSST~\citep{Gong2019}.} {Furthermore, the estimations at high redshift can be problematic without enough training data. To address these challenges, we supplement our slitless spectra utilizing data from the Baryon Oscillation Spectroscopic Survey (BOSS).} BOSS is a spectroscopic survey primarily targeting luminous red galaxies (LRGs) up to $z\sim0.7$ and quasars (QSOs) at redshifts $2.2<z<3$, aimed at detecting the characteristic scale imprinted by baryon acoustic oscillations (BAO) in the early Universe. Over its 5-year observation period, BOSS has measured spectra for approximately 4 million sources, covering 10,000 square degrees~\citep{Dawson2013}. For this work, we use 16th data release from BOSS (BOSS-DR16). 

Similar to our approach with DESI data, we select galaxies with a spectroscopic redshift warning ZWARN == 0, produced by Redrock software, and match these sources with the DESI LS DR9. We exclude sources modeled as PSF and without reasonable photometric measurements in the $g$, $r$ and $z$ bands and those lacking valid morphological parameters. This results in a selection of 450,000 galaxies, for which we download their model spectra using SPARCL. {Finally, we merge the sources from DESI and BOSS, obtaining approximately 600,000 sources in total.}

\begin{figure*}
    \centering
    \includegraphics[width=\textwidth]{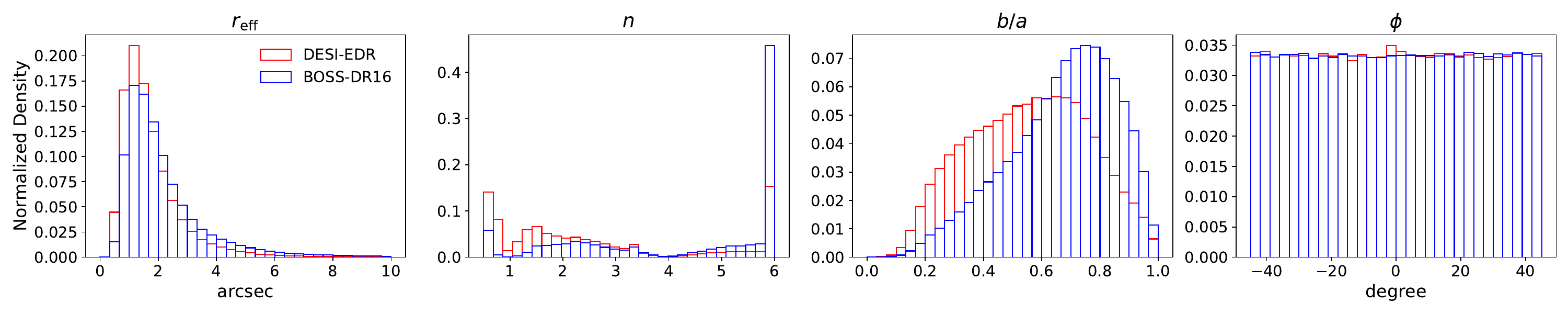}
    \caption{The distribution of four morphological parameters: effective radius $r_{\rm eff}$, sersic index $n$, axis ratio $b/a$ and position angle $\phi$ of sources from DESI-EDR and BOSS-DR16.}
    \label{fig:morph_params}
\end{figure*}

The spectroscopic redshift distributions are illustrated in Figure~\ref{fig:z_dist}. We notice that most sources from DESI-EDR are at low redshifts, while high redshift sources up to $z\sim1$ are supplemented by BOSS-DR16. The distributions of morphological parameters, including effective radius $r_{\rm eff}$, S{\'e}rsic index $n$, axis ratio $b/a$ and position angle $\phi$ are illustrated in Figure~\ref{fig:morph_params}. Notably, galaxies with a S{\'e}rsic index of 6 dominate, particularly those from BOSS-DR16. This selection bias is as expected, as the S{\'e}rsic index positively correlates with galaxy size and luminosity. {Hence, for valid morphological measurements and accurate spectroscopic redshift fitting, the sources are expected to be larger in size and brighter in luminosity, with large S{\'e}rsic indices.}

\begin{figure}
    \centering
    \includegraphics[width=0.5\textwidth]{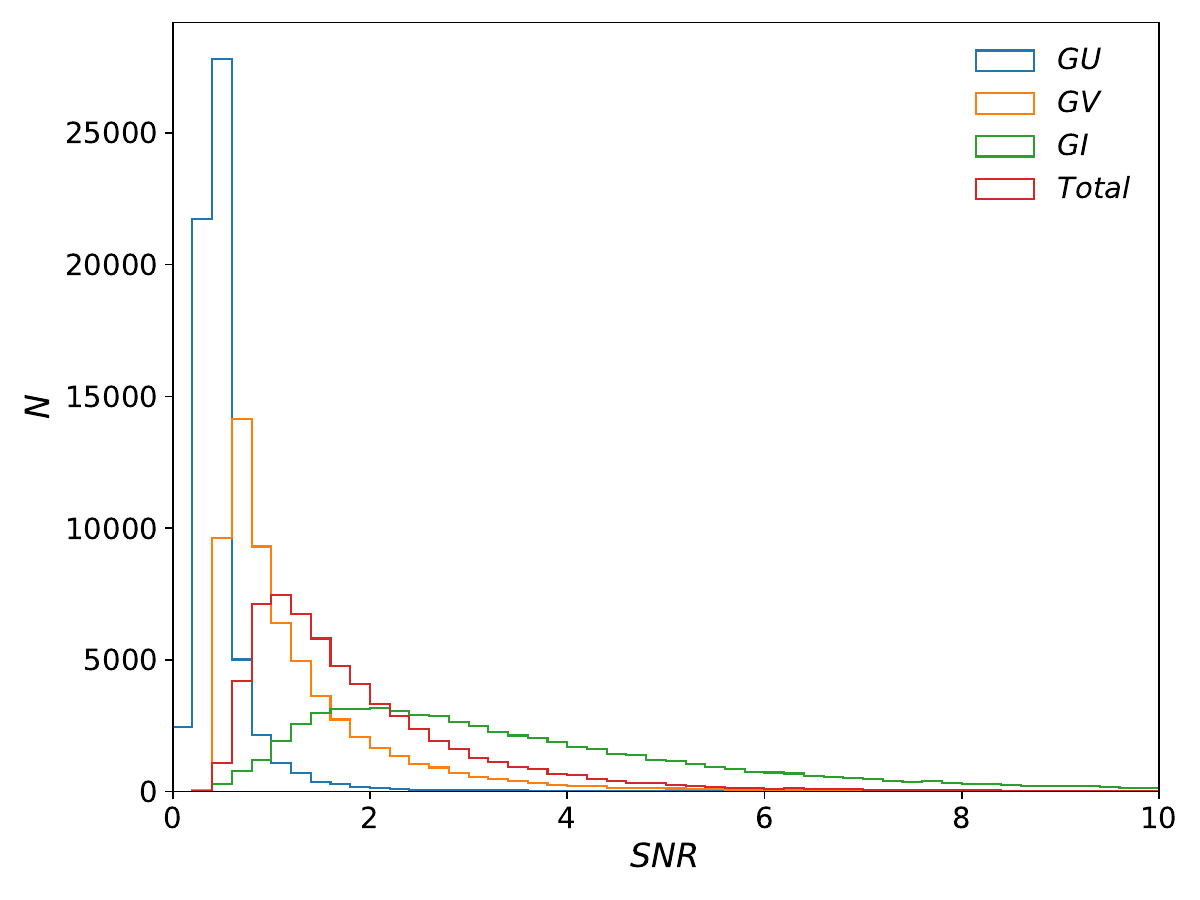}
    \caption{The SNR distributions of simulated slitless spectra in $GU$, $GV$, $GI$ bands and over the whole wavelength.}
    \label{fig:snr_dist}
\end{figure}

\begin{figure*}
    \centering
    \includegraphics[width=0.48\textwidth]{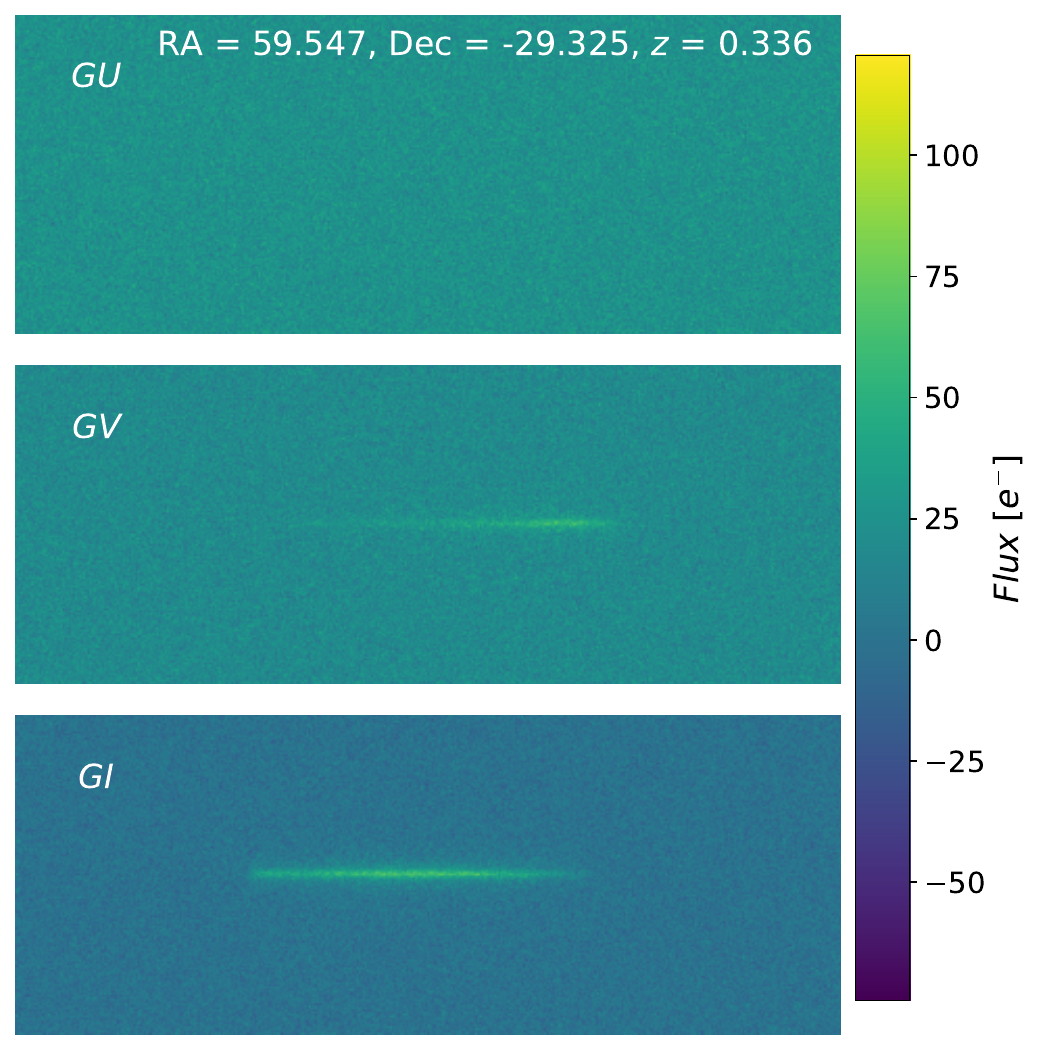}
    \vspace{0.5cm}
    \includegraphics[width=0.48\textwidth]{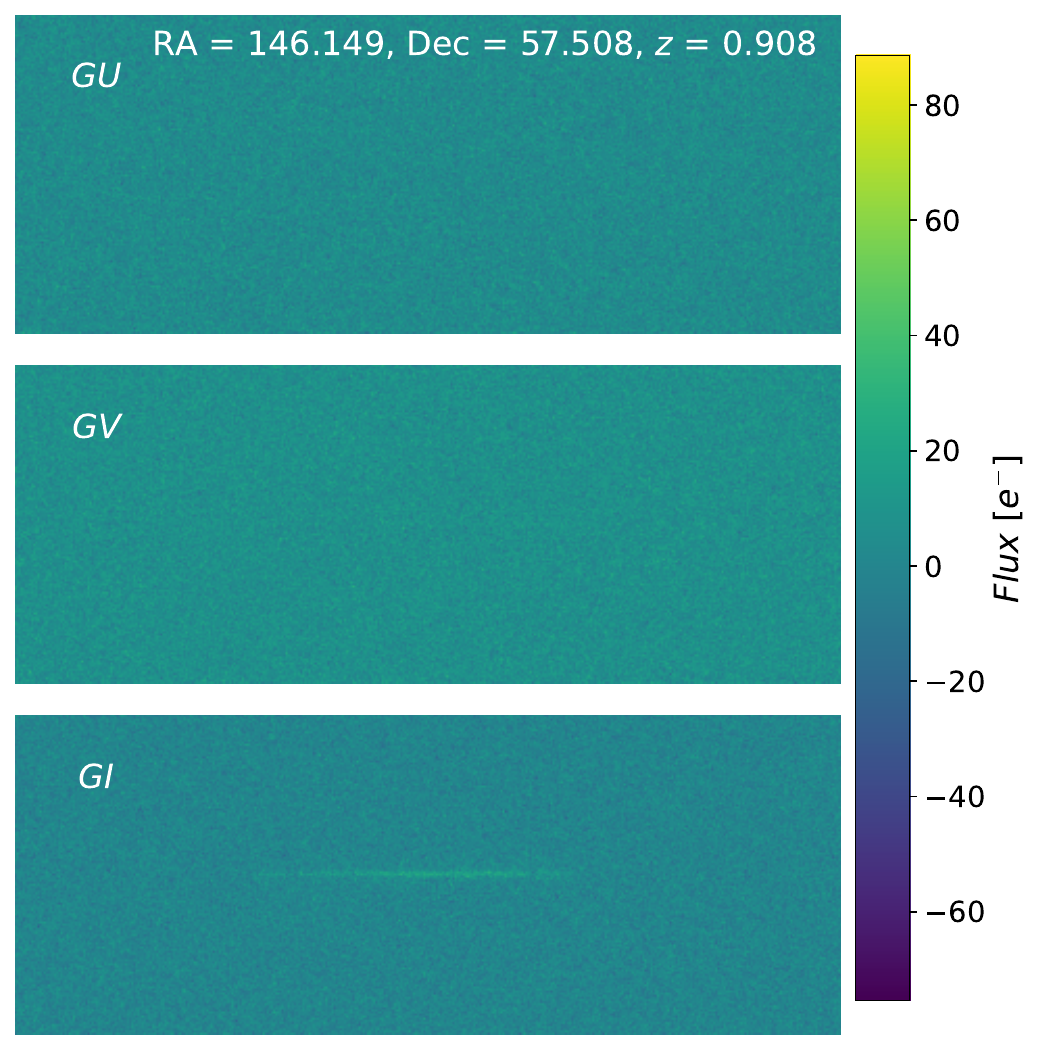}
    \caption{Two examples of simulated first-order slitless spectral images in $GU$, $GV$ and $GI$ bands. The coordinates and spectroscopic redshifts are also shown. And the corresponding extracted one-dimensional spectra and more information of the two sources are displayed in Figure~\ref{fig:sls_examples}.}
    \label{fig:spectra_2d_examples}
\end{figure*}

\begin{figure*}
    \centering
    \includegraphics[width=\textwidth]{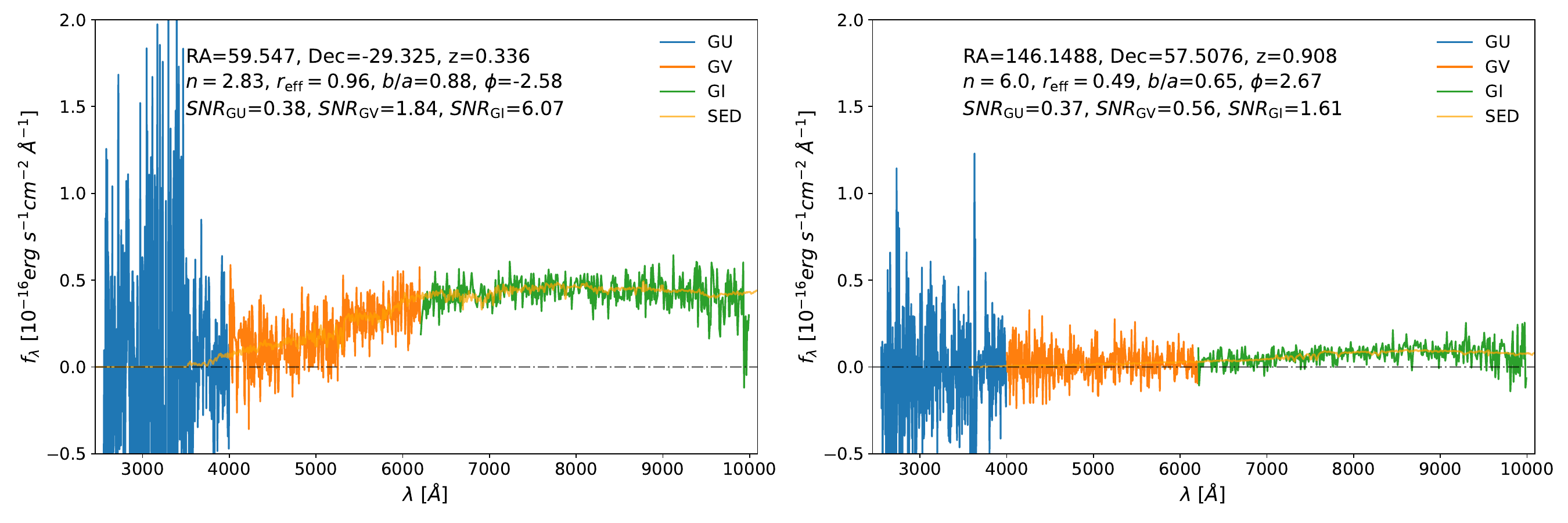} \\
    \caption{The corresponding one-dimensional spectra extracted from spectral images of sources in Figure~\ref{fig:spectra_2d_examples}. The SEDs used in simulation are also illustrated. And the black dash-dotted line indicates zero fluxes. Additionaly, the source information including coordinates, spectroscopic redshifts, morphological parameters and SNR of each band are also shown.}
    \label{fig:sls_examples}
\end{figure*}

After obtaining the model spectra and morphological parameters for sources in DESI-EDR and BOSS-DR16, we simulate the slitless spectra using the simulation software mentioned in Section~\ref{sec:software}. The signal-to-noise ratio (SNR) of the simulated CSST slitless spectra in $GU$, $GV$, $GI$ and total are illustrated in Figure~\ref{fig:snr_dist}. {We notice that the SNRs of $GI$ are the best reaching a peak at about $2$, while most SNRs of $GU$ and $GV$ bands are lower than $1$, especially for $GU$ band.} {The significantly low SNR in the $GU$ band can be attributed to the fact that the wavelength coverage of DESI and BOSS spectrograph does not fully encompass the wavelength coverage of the $GU$ band of CSST. Consequently, there are negative data points in the model spectra over the $GU$ band, and thus the slitless spectra in the $GU$ band are predominantly characterized by Gaussian noise.} Additionally, the total SNRs of these spectra peak at $\sim1$, indicating signal and noise are at similar level. In Figure~\ref{fig:spectra_2d_examples}, we display two examples of simulated first-order slitless spectral images in $GU$, $GV$ and $GI$ bands, and the corresponding extracted one-dimensional spectra are shown in Figure~\ref{fig:sls_examples}. The SEDs used in simulation are also illustrated and they are in consistency with the spectra. Additionally, the source information including coordinates (R.A. and Dec.), spectroscopic redshifts, morphological parameters used in simulation and SNR in $GU$, $GV$ and $GI$ band are also shown. For the low-redshift source in the left panel, we can clearly recognize the dispersed 2d-spectra in $GV$ and $GI$ bands, and SNRs of the extracted 1d-spectra in these two bands are relatively high. While for high-redshift source in the right panel, only a faint 2d-spectra in $GI$ band can be recognized, with the other two bands are dominated by noise, hence the extracted 1d-spectra are correspondingly with low SNR. Overall, the slitless spectra are severely affected by background and instrumental noise and the recognition of spectroscopic features such as break, absorption and emission lines is difficult, leading to challenges for successful redshift determinations using traditional approaches such as spectrum fitting or feature identification. 

\section{Methodology} \label{sec:methods}
{We employ convolutional neural network (CNN) to estimate redshifts from slitless spectra expected to be observed by CSST. To satisfy the requirement of some cosmological studies, we further construct Bayesian neural network (BNN) to derive redshift values along with their uncertainties. Our networks are constructed using Keras~\footnote{{\url{https://keras.io/}}} backend on TensorFlow~\footnote{\url{https://www.tensorflow.org/}} and TensorFlow-Probability~\footnote{\url{https://www.tensorflow.org/probability}}, and the relevant codes are publicly available at Github~\footnote{\url{https://github.com/xczhou-astro/CSST_slitless_spectra}}.}

\subsection{Neural networks}\label{sec:neural_network}

\begin{figure*}
    \centering
    \includegraphics[width=0.70\textwidth]{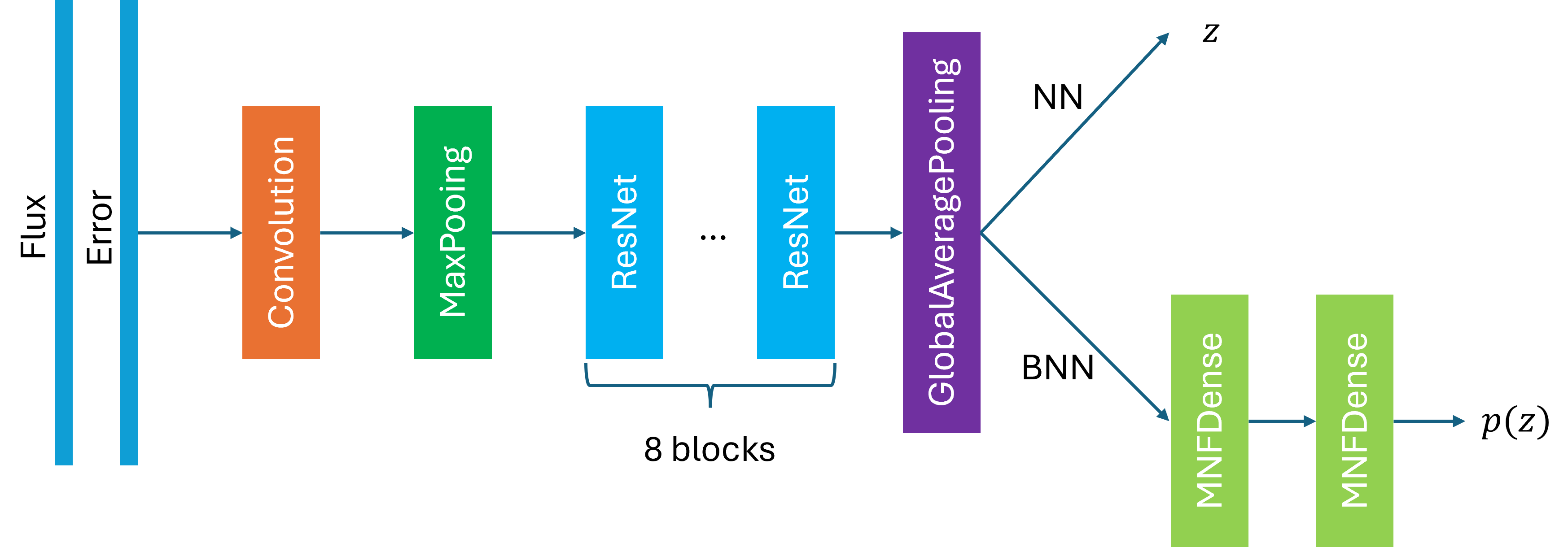}
    \hspace{1cm}
    \includegraphics[width=0.2\textwidth]{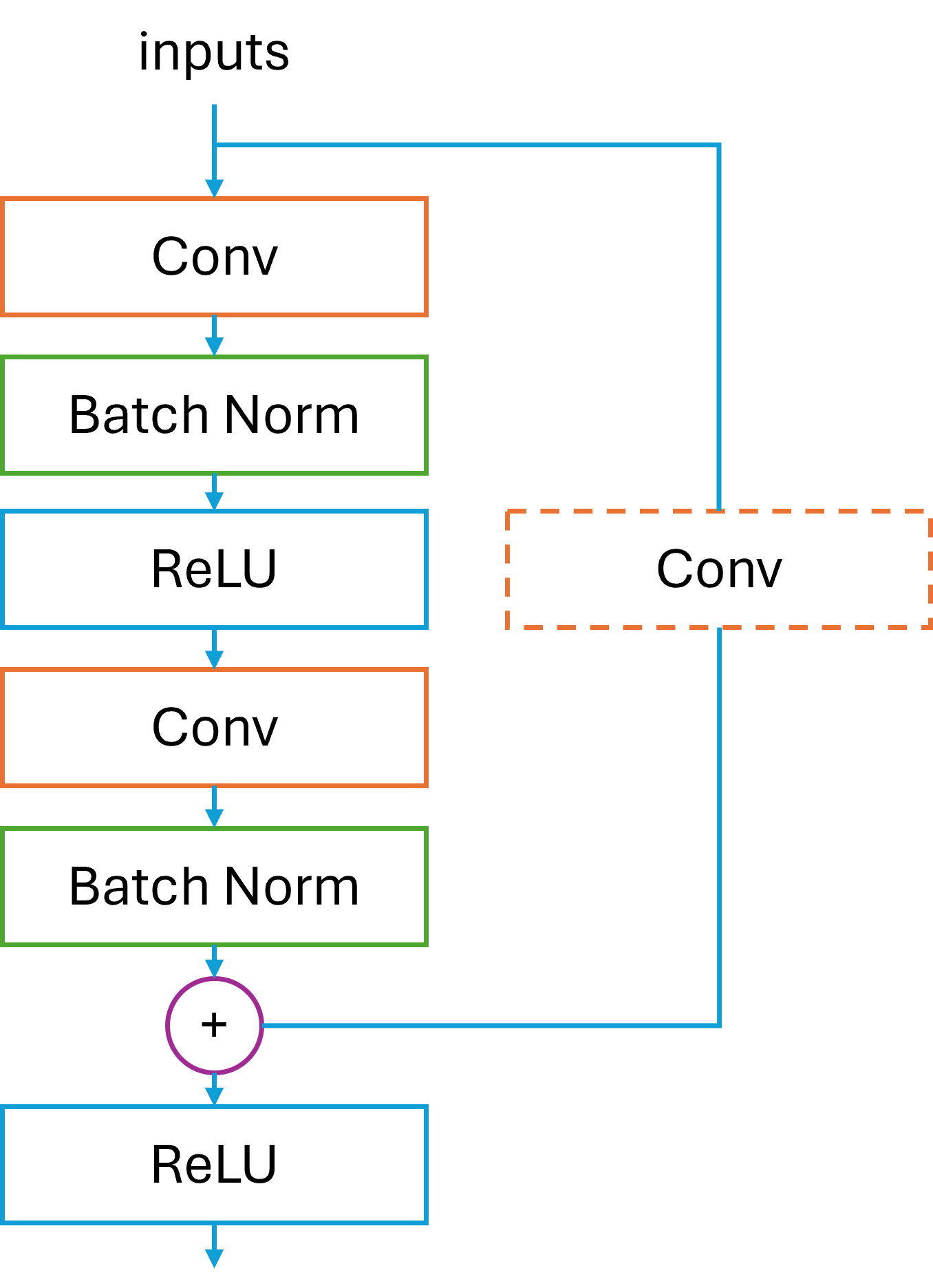}
    \caption{\textit{Left:} The architecture of the 1d-CNN and BNN built upon ResNet blocks. 
    \textit{Right:} The structure of the ResNet block.}
    \label{fig:architecture}
\end{figure*}

Since spectra are 1-dimensional sequences, we employ 1d-CNN to process them. CNN is a powerful deep learning model which can learn the internal connections between data and labels. Therefore, we expect that our 1d-CNN can learn the mapping between slitless spectra and redshifts. To improve its learning ability, we increase the depth of our 1d-CNN using ResNet blocks~\citep{He2015}. This block can effectively reduce the vanishing gradient problem commonly happened in deep neural networks through skip connections, as illustrated in the right panel of Figure~\ref{fig:architecture}. The convolutional layer in skip connection is applied when this block process and downsample the features at the same time. Following~\citet{Zhou2021}, the input to our CNN includes spectra and corresponding errors as a two-channel sequences. And then the inputs are processed by one convolutional layer with 32 kernels with kernel size of 7, followed by a max-pooling to reduce the feature dimension. After this shallow feature extraction layers, we structure 8 ResNet blocks to obtain useful features from spectra and further reduce the feature dimension. The 1d convolutional layers in these blocks are all followed by BatchNormalization layers~\citep{Ioffe2015} to reduce overfitting. And ReLU activation functions~\citep{Agarap2018} are structured to apply non-linearity. Subsequently, the features are vectorized to one-dimension using global average pooling, and then a dropout layer~\citep{Srivastava2014} with drop rate of 0.2 is applied. This dropout layer also functions as overfitting reduction. Finally, the output layer with one neuron is structured. The illustration of architecture is displayed in the left panel of Figure~\ref{fig:architecture}. 

The neural network discussed yet can only output redshift values. {To output redshifts along with their uncertainties, we construct a Bayesian neural network (BNN). The uncertainties can be categorized into two types: epistemic uncertainty, which arises from neural network models and can be reduced by averaging ensemble networks with different configurations; and aleatoric uncertainty, which originates from intrinsic corruption of the data and cannot be reduced~\citep{Kiureghian2009,Hullermeier2019}. BNN can capture both uncertainties by varying the network configurations and utilizing a specific distribution as the output, with the Gaussian distribution being the most common choice.} For more detailed discussion on this network, please refer to~\citet{Hortua2020} and ~\citet{Zhou2022b}. Our BNN is built using transfer learning technique to transfer the feature extraction part before the final output layer of CNN and then append two Bayesian layers. The weights from the transferred network are set as fixed, leveraging the features that are tailored to derive the redshift. For Bayesian layers, three common structures are commonly employed, i.e. Monte-Carlo dropout (MC-dropout, ~\cite{Gal2015}), flipout~\citep{Wen2018}, and Multiplicative Normalizing Flows (MNF, ~\citet{Louizos2017}) layer. Among these three categories, MC-dropout uses dropout to simulate varying configurations of network, while the weights in flipout and MNF layers are represented by distributions. Particularly, MNF employs more complicated distributions transformed from Gaussian distribution using normalizing flows~\citep{Jimenez2015}. As recommended by another work (Zhou et al. in preparation), we adopt MNFDense layers~\footnote{\url{https://github.com/janosh/tf-mnf}} adapting 50 layers for masked RealNVP normalizing flow~\citep{Dinh2016}. {Since Gaussian distribution is a reasonable assumption for redshift estimation, the network finally outputs two values, from which a Gaussian distribution can be derived.} The illustration of architecture is also displayed in the left panel of Figure~\ref{fig:architecture}.

{Bayesian networks inherently introduce uncertainties that are either overestimated or underestimated, deviating from the statistical principle that the coverage probabilities of the sample with the true value within specific confidence intervals correspond to the corresponding confidence intervals~\citep{Perrault2017, Hortua2020}. To assess the calibration status, we can employ the reliability diagram, which plots the coverage probability against the confidence interval. The uncertainties are well-calibrated when this diagram exhibits a straight diagonal line. Conversely, calibration is essential before reporting the results. Extensive discussions on calibration techniques can be found in~\citet{Guo2017, Hortua2020}. In this study, we adopt a straightforward after-training calibration technique, Beta calibration~\citep{Kull2017}, to calibrate the uncertainties. This calibration procedure ensures that the uncertainties more accurately reflect the true confidence intervals, thereby enhancing the reliability of our redshift estimations.} 

\subsection{Training}\label{sec:training}
{We only consider the spectra in $GV$ and $GI$ bands, since the SNRs in $GU$ band are significantly low and the spectra are dominated by noise as explained in Section~\ref{sec: data generation}.} Before training, our data are split into training, validation and testing sets as a ratio of 8:1:1~\footnote{Data are available at \url{https://pan.cstcloud.cn/s/E6FrFGa6TJA}}. The number of testing set is approximately 60,000 and they are selected based on the expected redshift distribution of CSST slitless spectroscopic survey. {The redshift distribution of the testing set is displayed in Figure~\ref{fig:z_dist_test}, and is consistent with the one shown in~\citet{Gong2019} at low redshift region.} To improve the performance of neural network, we follow the methodology in~\citet{Zhou2021} to increase the size of training set by involving their Gaussian realization counterparts created through fluctuating the spectra based on the their corresponding errors. This data augmentation technique can effectively amplify the adaptability of network to the large noise in low-SNR siltless spectra. Here we use 50 random realizations. For 1d-CNN, we set loss function and optimizer as logcosh and Adam. Logcosh resembles traditional mean-absolute-error function but with a differential behavior around 0, and Adam is a stochastic gradient descent optimization method based on adaptive estimate of first-order and second-order moments~\citep{Kingma2014}. This network is trained for 100 epochs with batch size of 1024 considering the graphic memory of GPU. {We select the best model with lowest loss value as our final CNN model and the backbone for BNN, since the performance of CNN backbone substantially influences the outcomes of the BNN.}

For BNN, only the weights of appended MNF layers are optimized in training. {This training strategy can leverage the features optimized for estimating redshifts and substantially increase the training speed.} The loss function of BNN is set to be negative log-likelihood (NLL), different from the one employed in CNN, since BNN outputs a distribution considering both point values and their uncertainties. Note that the labels are solely redshift values, since the uncertainties are naturally derived during the decrease of the loss function. Similarly, we adopt the Adam optimizer and save the model with lowest loss value. Different from the CNN in post processing, we feed the testing spectra to the BNN for 200 times. Based on these outputs, we calculate the final redshift values and their corresponding uncertainties which account for both epistemic and aleatoric ones. 

{The main hyperparameters for CNN and BNN we tune and adopt are summarized in Table~\ref{tab:hyperparameters}. Note that the hyperparameters do not severely affect the performance for both networks as long as the network is deep enough and the training data is sufficient.}

\begin{table}

\caption{Main hyperparameters that we tune and adopt for the CNN and BNN.}
\label{tab:hyperparameters}
\begin{center}
\begin{tabular}{lcc}
\hline
                          & CNN  & BNN  \\ \hline \hline
Kernel size of 1d-CNN     & 7    & -$^a$    \\ \hline
Dropout rate              & 0.2  & -    \\ \hline
Number of ResNet blocks   & 8    & -    \\ \hline
Number of Bayesian layers & -    & 2    \\ \hline
Learning rate for Adam             & 1e-4 & 1e-4 \\ \hline
Loss function             & LogCosh$^b$ & NLL \\ \hline
Batch size                & 1024 & 1024 \\ \hline
Number of augments        & 50   & 50   \\ \hline
\end{tabular}
\end{center}
\textbf{Notes}.\\
$^a$ Inherited from CNN through transfer learning.\\
$^b$ LogCosh and HuberLoss are experimented.\\
\end{table}

\section{Results}\label{sec:results}
We employ two metrics to evaluate the performance of CNN: outlier percentage $\eta$ and normalized median absolute deviation $\sigma_{\rm NMAD}$, defined as follows:
\begin{equation}
    \eta = \frac{N_{\Delta z/(1 + z_{\rm true}) > 0.02}}{N_{\rm total}},
\end{equation}
\begin{equation}
    \sigma_{\rm NMAD} = 1.48\times{\rm median}\left(\left|\frac{\Delta z - {\rm median}(\Delta z)}{1 + z_{\rm true}} \right|\right),
\end{equation}
where $\Delta z = z_{\rm pred} - z_{\rm true}$, with $z_{\rm pred}$ and $z_{\rm true}$ indicating the predictions and true redshifts respectively. {$\eta$ demonstrates the fraction of severely inaccurate redshift predictions, and $\sigma_{\rm NMAD}$ is an accuracy metric that is robust against the outliers.}

\begin{figure}
    \centering
    \includegraphics[width=0.5\textwidth]{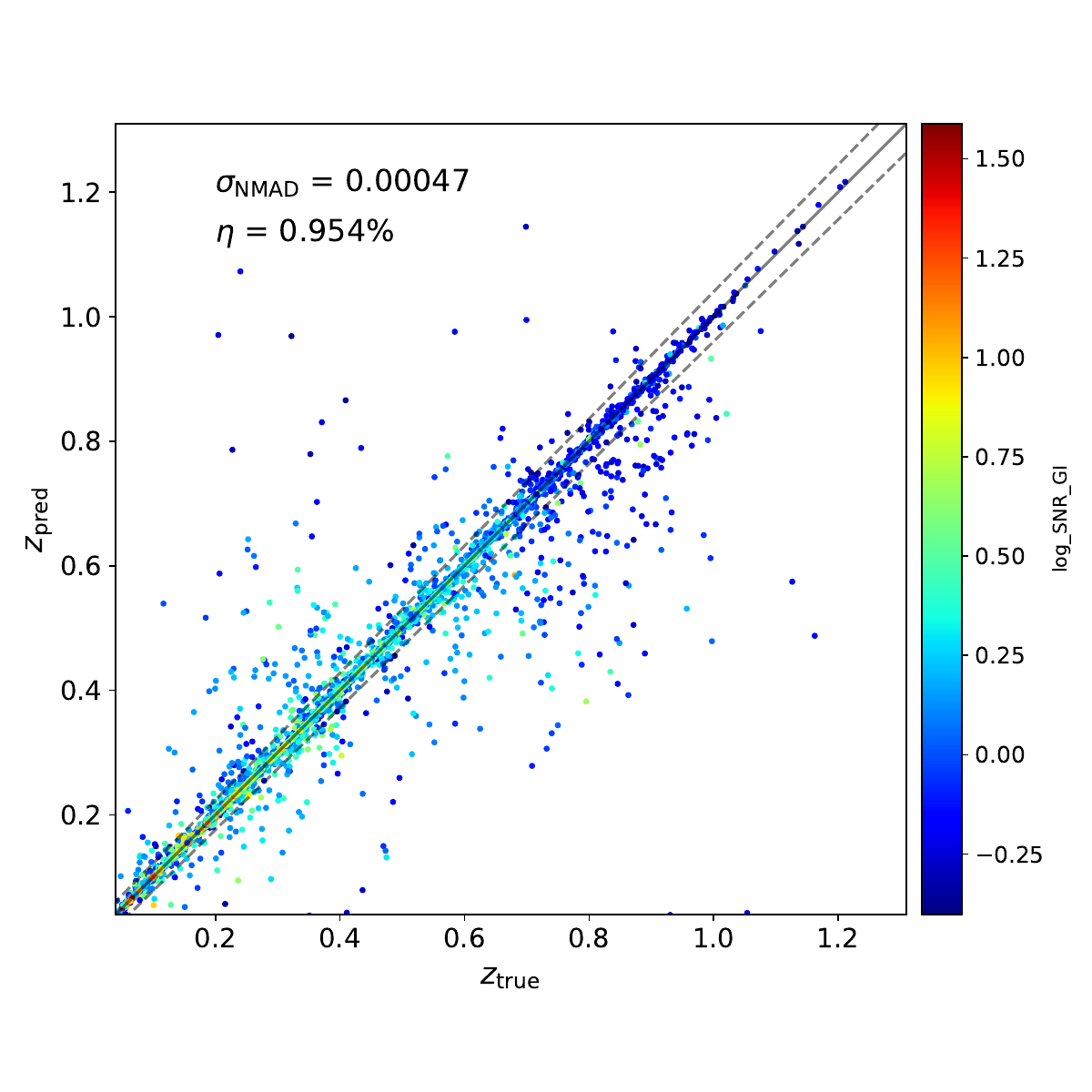}
    \caption{The results of 1d-CNN are illustrated, achieving accuracy $\sigma_{\rm NMAD}=0.00047$ and outlier percentage $\eta=0.954\%$ respectively. The logarithmic SNR of $GI$ bands are indicated by colorbar. }
    \label{fig:points}
\end{figure}

\begin{figure}
    \centering
    \includegraphics[width=0.5\textwidth]{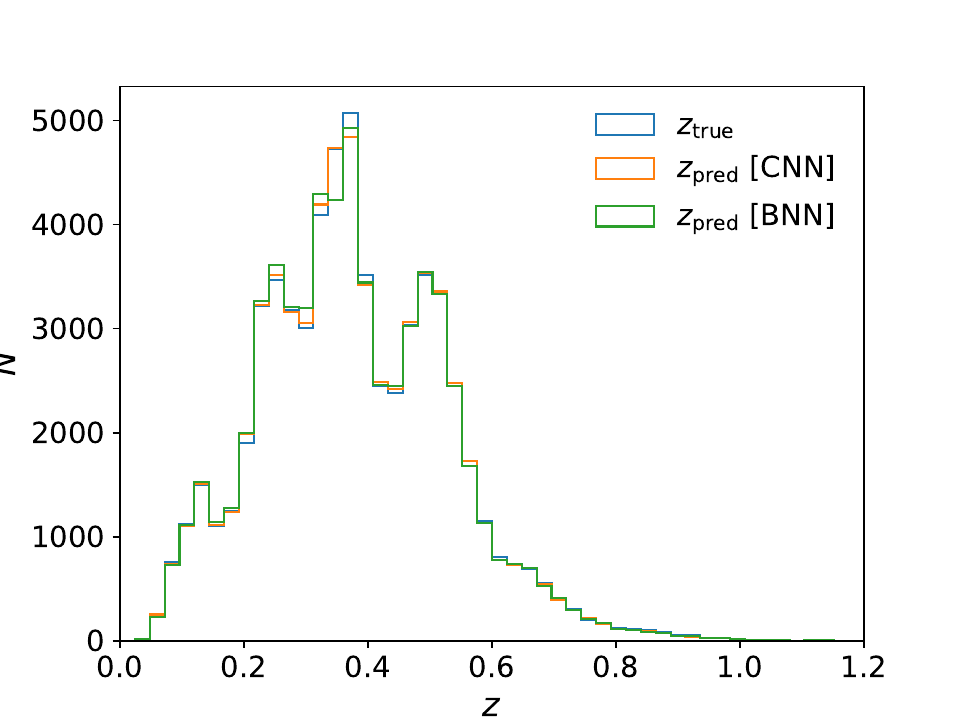}
    \caption{The distributions of true and predicted redshifts from CNN and BNN for testing data.}
    \label{fig:z_dist_test}
\end{figure}

\begin{figure}
    \centering
    \includegraphics[width=0.5\textwidth]{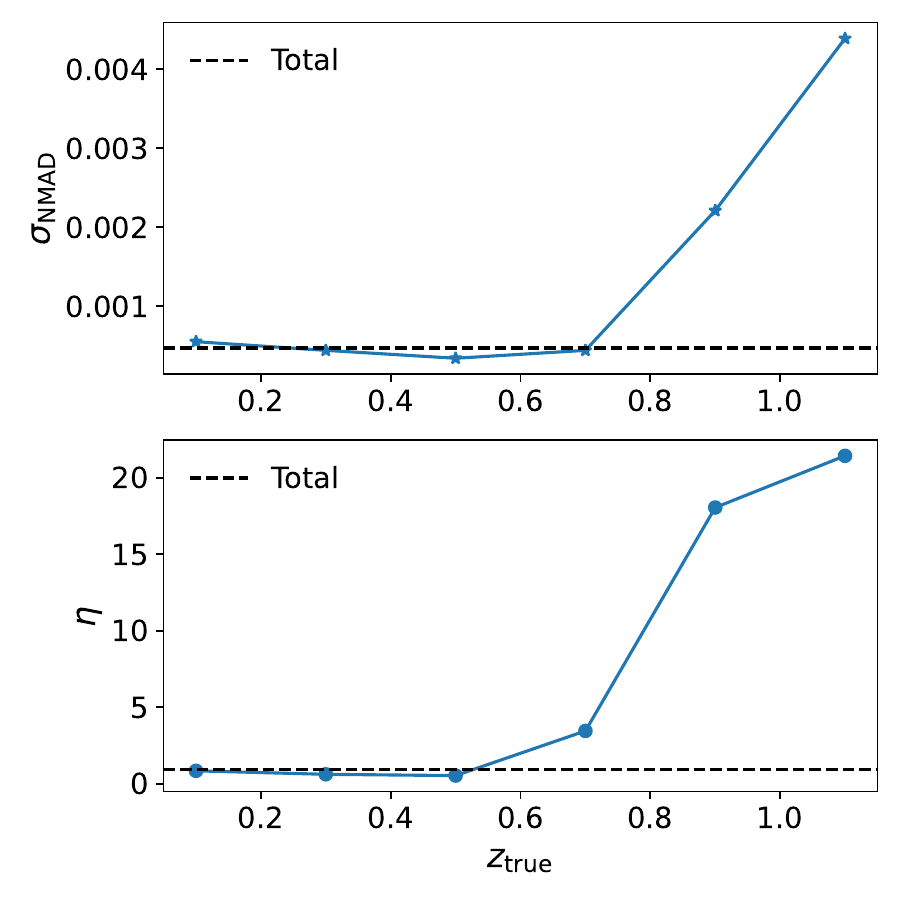}
    \caption{The accuracy $\sigma_{\rm NMAD}$ and outlier percentage $\eta$ with respect to true redshifts are displayed in the upper and lower panel respectively. The two metrics over whole redshift range are also shown in black dashed lines.}
    \label{fig:stats_point}
\end{figure}

Figure~\ref{fig:points} illustrates the results for CNN. The accuracy $\sigma_{\rm NMAD}$ and outlier percentage $\eta$ can reach 0.00047 and 0.954\% respectively. {The accuracy successfully fulfills the requirement of $\sigma_{\rm NMAD} < 0.005$ for BAO and other studies employing CSST slitless spectroscopic surveys~\citep{Miao2024, Gong2019}.} The logarithmic SNR of $GI$ bands are also displayed by colorbar. We notice that as expected, the SNR decreases with respect to redshift and most outliers have relatively low SNR. The redshift distribution is displayed in Figure~\ref{fig:z_dist_test}, and is in high consistency with true distribution. Furthermore, Figure~\ref{fig:stats_point} displays the accuracy and outlier percentage with respect to true redshifts in the upper and lower panel respectively. The two metrics over whole redshift range are also displayed in black dashed lines. As expected, both metrics remain steady at lower redshift and becomes worse at higher redshift. 

\begin{figure}
    \centering
    \includegraphics[width=0.5\textwidth]{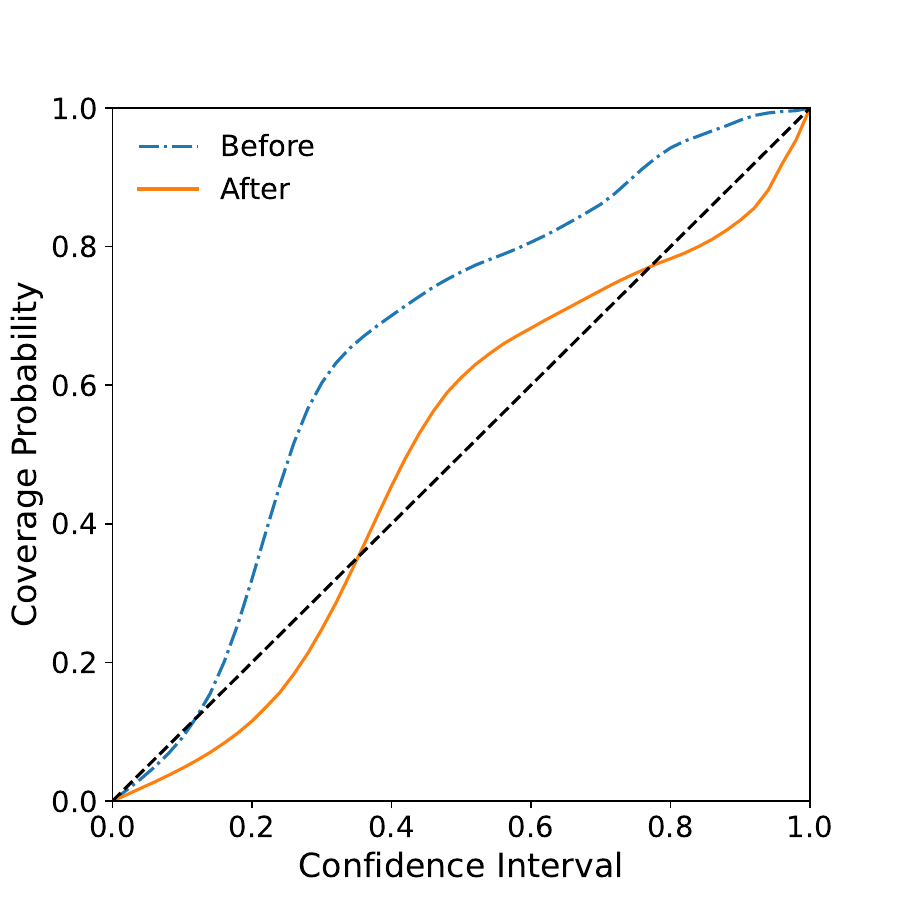}
    \caption{Reliability diagram for BNN results before and after the calibration. The black dashed line indicates that the uncertainties are well-calibrated with statistical principle perfectly followed.}
    \label{fig:qq_plot}
\end{figure}

\begin{figure}
    \centering
    \includegraphics[width=0.5\textwidth]{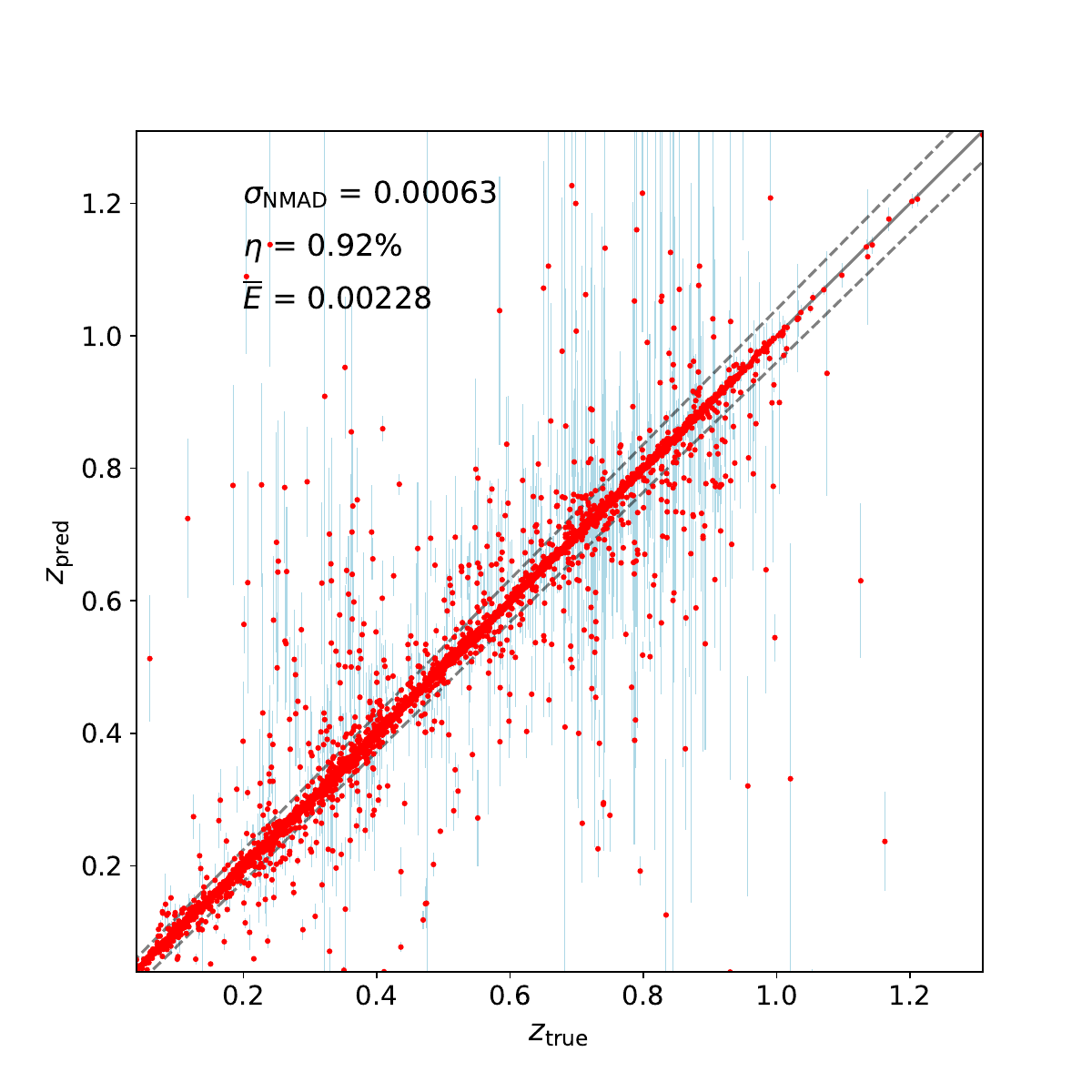}
    \caption{The results of BNN after uncertainty calibration. The errorbars are displayed in lightblue. Over the whole redshift range, BNN can reach $\sigma_{\rm NMAD}=0.00063$ and $\eta=0.92\%$ respectively. And weighted mean uncertainty $\overline{E}$ can achieve 0.00228.}
    \label{fig:BNN_results}
\end{figure}

\begin{figure}
    \centering
    \includegraphics[width=0.5\textwidth]{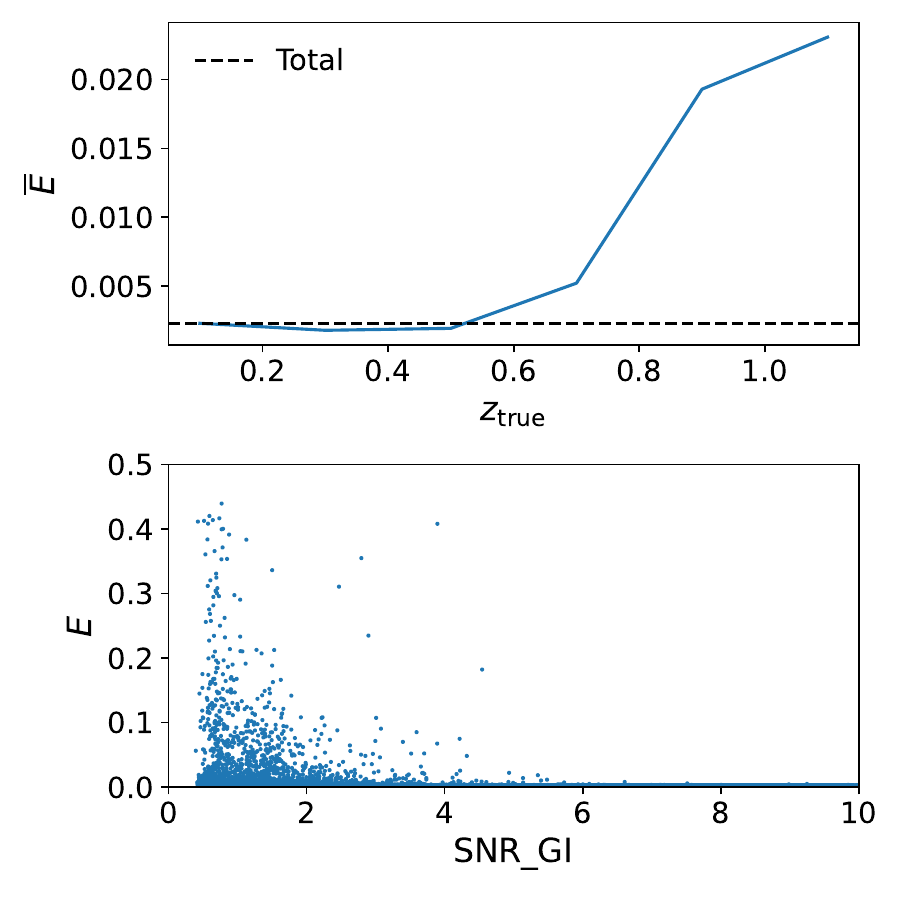}
    \caption{\textit{Upper:} weighted mean uncertainty $\overline{E}$ with respect to true redshift. The value over whole redshift range is also displayed in black dashed line. \textit{Lower:} weighted uncertainty $E$ with respect to SNR in $GI$ band.}
    \label{fig:error_analysis}
\end{figure}

As for the results from BNN, apart from the two metrics mentioned above, we employ another metric to measure the performance of uncertainty predictions, i.e. weighted mean uncertainty $\overline{E}$, which is defined as:
\begin{equation}
    \label{eq:weighted_mean_uncertainty}
    \overline{E} = \frac{\sum_i E_i/(1 + z_{i, \rm{true}})}{N_{\rm total}}, 
\end{equation}
where $E_i$ is the uncertainty prediction for each source. The weight $1 + z_{i,\rm true}$ applied to each source is to eliminate the bias from the evolution of redshifts. {Figure~\ref{fig:qq_plot} displays the reliability diagram for uncertainty predictions, from which we notice that the uncertainties are overestimated.} After Beta calibration, the uncertainties better follow the statistical principle as mentioned in Section~\ref{sec:neural_network}. And Figure~\ref{fig:BNN_results} shows the results after uncertainty calibration, where the errorbars are displayed in lightblue. $\sigma_{\rm NMAD}$ and $\eta$ can achieve 0.00063 and 0.92\% respectively. $\eta$ slightly improves compared to point estimates illustrated in Figure~\ref{fig:points}, while the accuracy becomes a little worse, but still satisfy the requirement of cosmological studies. Furthermore, the weighed mean uncertainty $\overline{E}$ can reach 0.00228. The redshift distribution is displayed in Figure~\ref{fig:z_dist_test}, and similarly is consistent with true distribution. Figure~\ref{fig:error_analysis} further analyze the behavior of uncertainties. The upper panel displays the weighted mean uncertainty $\overline{E}$ with respect to true redshift, in which the black dashed line shows the value over the whole redshift range. As expected, this metric similarly remains stable at lower redshift and becomes worse as redshift increases. {The lower panel shows the scatter plot between uncertainty and SNR in $GI$ band, and we notice that with SNR increasing, the scatter of uncertainties decreases as expected.}

{For comparison, according to the data analysis pipeline for CSST slitless spectra, the traditional fitting for redshift estimations may produce an accuracy as low as $\sigma_{\rm NMAD}\sim0.01$ under such low SNR $\sim1$ shown in Figure~\ref{fig:snr_dist}.} This demonstrates that deep learning algorithm can significantly enhance the accuracy of redshift estimations, especially for low-SNR slitless spectra.

\section{Discussion}\label{sec:discussion}
{As outlined in Section~\ref{sec:software}, the slitless spectra simulation software accepts SEDs and morphological parameters. The morphological parameters are employed to generate a S{\'e}rsic profile for each galaxy. This profile serves as the basis for dispersion. Nevertheless, not all galaxy have valid measurements of morphological parameters. Therefore, our selected sources are predominantly biased toward bright galaxy sample (BGS) and luminous red galaxies (LRG) at low redshift, with relatively larger and brighter luminosity. Emission line galaxies (ELGs), an ideal tracer at higher redshift reaching $\sim1.6$, are unfortunately excluded due to inadequate morphological measurements. Consequently, the selected samples are unable to achieve the redshift limit of $\sim1.5$ anticipated for the CSST spectroscopic survey, thereby hindering the comprehensive investigations of redshift accuracy by deep learning method.  Additionally, not all galaxies can be adequately represented by S{\'e}risc profiles, particularly considering the siginficant uncertainties associated with some morphological measurements. Furthermore, the morphologies of some galaxies may exhibit significant variations across different spectroscopic bands, rendering a single S{\'e}rsic profile inadequate. These biases and challenges can potentially be mitigated and addressed by incorporating a novel feature that directly simulates slitless spectra with galaxy images in reference photometric bands that are close to the slitless bands, like F140W to G141 and F098 to G102 in HST~\citep{Marinelli2024}. }

{On the other hand, the analysis of slitless spectra also presents challenges, particularly in terms of wavelength calibration and blending effects. These challenges severely impact the estimation of redshifts for individual galaxies. In the context of the CSST, wavelength calibrations are more complex compared to HST~\citep{Marinelli2024} due to the absence of a direct image as a reference for determining wavelengths, considering the focal plane arrangement of CSST. This can be addressed by employing convolutional neural networks (CNNs) to directly estimate redshifts from 2D spectral images. CNNs can analyze spectral images without wavelength information by leveraging knowledge gained from extensive training data. And blending effects are prevalent in spectroscopic observations, especially in deep surveys. Deblending algorithms, whether traditional or deep learning-based, are actively being researched~\citep{Burke2019,Arcelin2021,Hemmati2022} and will be thoroughly investigated for slitless spectra in the future.}


\section{Conclusion}\label{sec:conclusion}
In this work, we employ neural network to estimate the redshift from simulated slitless spectra in the CSST slitless spectroscopic survey. The simulation requires SED and four morphological parameters including effective radius, sersic index, axis ratio and position angle of each galaxy. To simulate the slitless spectra realistically, we use observational data from DESI-EDR and BOSS-DR16 with high-quality spectroscopic redshifts. The model spectra generated in spectrum fitting process for these two observations are considered as SEDs, and the sources are matched with DESI LS DR9 to retrieve the required morphological parameters. The SNR of the slitless spectra are low with total SNRs peaking at $\sim1$, hence the key spectroscopic features used for redshift determinations are hard to identify. Therefore, we leverage the superior capability in processing noisy data of neural network to estimate the redshifts from these slitless spectra.

Recognizing the importance of uncertainty predictions for several cosmological studies, we employ Bayesian network to accomplish this task by providing redshift estimations along with uncertainties. To increase the robustness and converging speed, we construct the BNN based on a CNN for point estimates using transfer learning techniques. Gaussian random realizations are employed to largely augment the training size, ensuring the generative ability and noise tolerance of BNN. After training, the uncertainty predictions for testing data are carefully calibrated. {The BNN can achieve the results of $\sigma_{\rm NMAD} = 0.00063$, $\eta=0.92\%$ and $\overline{E}=0.00228$, successfully satisfying the performance requirement of accuracy $\sigma_{\rm NMAD} < 0.005$ for BAO and other cosmological studies based on slitless spectra of CSST.} Our approach can achieve better performance than traditional SED fitting, particularly for low SNR slitless spectra, serving as a complementary method for spectroscopic redshift estimation.

{We also recognize that our work has certain limitations. These limitations can be attributed to our current version of simulation software, and the limited data size from DESI early data release. With newer version of software and future data releases from DESI, we anticipate a comprehensive investigation of the spectroscopic redshift accuracy that can be achieved by deep learning algorithms for CSST slitless spectroscopic survey. }

\begin{acknowledgments}
XCZ and YG acknowledge the support from National Key R\&D Program of China grant 2022YFF0503404, 2020SKA0110402, and the CAS Project for Young Scientists in Basic Research (no. YSBR- 092). This work is also supported by science research grants from the China Manned Space Project with grant nos. CMS-CSST-2021-B01 and CMS-CSST-2021-A01. NL acknowledge the support from The science research grants from the China Manned Space Project (No. CMS-CSST-2021-A01), The CAS Project for Young Scientists in Basic Research (No. YSBR-062) and The Ministry of Science and Technology of China (No. 2020SKA0110100)
HZ acknowledges the science research grants from the China Manned Space Project with Nos. CMS-CSST-2021-A02 and CMS-CSST-2021-A04 and the supports from National Natural Science Foundation of China (NSFC; grant Nos. 12120101003 and 12373010) and  National Key R\&D Program of China (grant Nos. 2023YFA1607800, 2022YFA1602902) and Strategic Priority Research Program of the Chinese Academy of Science (Grant Nos. XDB0550100).
\end{acknowledgments}

\bibliography{sample631}{}
\bibliographystyle{aasjournal}



\end{document}